\documentclass[conference]{IEEEtran}
\ifCLASSINFOpdf
\else
\fi
\usepackage{graphicx}
\usepackage{setspace}
\usepackage{amsmath,empheq}
\usepackage{amssymb}
\usepackage{amsthm}
\usepackage{caption}
\usepackage{subcaption}
\usepackage{bm} 
\usepackage{cite}
\usepackage{float}
\usepackage[usenames,dvipsnames]{pstricks}
\usepackage{epsfig}
\usepackage{pst-grad} 
\usepackage{pst-plot} 
\usepackage{tabularx}
\usepackage[keeplastbox]{flushend}
\usepackage{amsmath}
\usepackage[T1]{fontenc}
\newcommand{\lb} {\left}
\newcommand{\rb} {\right}
\newcommand{\nn} {\nonumber}
\usepackage{xcolor}
\usepackage{lipsum}
\usepackage{comment}

\allowdisplaybreaks

\allowdisplaybreaks
\flushend
\usepackage{caption}
\usepackage[utf8]{inputenc}

 \IEEEaftertitletext{\vspace{-2.5\baselineskip}}
\begin{document}
\onecolumn{\noindent © 2022 IEEE. Personal use of this material is permitted. Permission from IEEE must be obtained for all other uses, in any current or future media, including reprinting/republishing this material for advertising or promotional purposes, creating new collective works, for resale or redistribution to servers or lists, or reuse of any copyrighted component of this work in other works.}
 \twocolumn{
\title{Optimal Friendly Jamming and Transmit Power Allocation in RIS-assisted Secure Communication}

\author{
    \IEEEauthorblockN{Burhan Wafai\IEEEauthorrefmark{1}, Chinmoy~Kundu\IEEEauthorrefmark{2}, 
    Ankit Dubey\IEEEauthorrefmark{3},
    and Mark F. Flanagan\IEEEauthorrefmark{4},
    }
    \IEEEauthorblockA{\IEEEauthorrefmark{1}\IEEEauthorrefmark{3}Department of EE, Indian Institute of Technology Jammu, Jammu \& Kashmir, India }
     \IEEEauthorblockA{\IEEEauthorrefmark{2}\IEEEauthorrefmark{4}School of Electrical and Electronic Engineering, University College Dublin, Belfield, Ireland}
    \textrm{\{\IEEEauthorrefmark{1}burhan.wafai}, {\IEEEauthorrefmark{3}ankit.dubey}\}@iitjammu.ac.in,{\IEEEauthorrefmark{2}chinmoy.kundu@ucd.ie},
     {\IEEEauthorrefmark{4}mark.flanagan@ieee.org}
     }

\maketitle
\begin{abstract}
This paper analyzes the secrecy performance of a reconfigurable intelligent surface (RIS) assisted wireless communication system with a friendly jammer in the presence of an eavesdropper. The friendly jammer enhances the secrecy by introducing artificial noise towards the eavesdropper without degrading the reception at the destination. Approximate secrecy outage probability (SOP) is derived in closed form. We also provide a simpler approximate closed-form expression for the SOP in order to understand the effect of system parameters on the performance and to find the optimal power allocation for the transmitter and jammer. The optimal transmit and jamming power allocation factor is derived by minimizing the SOP assuming a total power constraint. It is shown that the SOP performance is significantly improved by the introduction of the jammer and a gain of approximately $3$ dB is achieved at an SOP of $10^{-4}$ by optimally allocating power compared to the case of equal power allocation. 
\end{abstract}

\section{Introduction}

To address the demands of next-generation communication networks, reconfigurable intelligent surface (RIS) technology 
is a promising candidate for enabling smart control of the propagation of
signals. An RIS consists of a large number of passive meta-material elements, and can intelligently control the phase shift and amplitude at each element and is also energy efficient due to its passive elements \cite{Zhang_towards_ris,renzo_wireless_com_RIS}. 

Despite RIS being a relatively new technology, its integration with other wireless technologies has been studied extensively. Enhanced energy detection using RIS was 
proposed in \cite{Wu_cognitive_ris} 
for single-user spectrum sensing, cooperative spectrum sensing; the performance of diversity reception and average probability of detection was studied in each case.  
The authors in \cite{Xu_user_selection_ris} studied the ergodic capacity of user scheduling in RIS-assisted  multi-user communication.  
In \cite{RIS_Amarasuriya}, the authors analyzed the performance of an RIS-aided communication system in terms of the outage probability,
average symbol error probability, and
achievable rate bounds.
Although the performance analysis in an RIS-assisted communication is studied in the aforementioned papers, the secrecy aspect has not been considered. 

The broadcast nature of wireless transmission is always vulnerable to eavesdropping.  Hence the security of such networks is of utmost importance, and consequently,  physical layer security (PLS) has emerged as a viable solution \cite{wyner_wiretap}. 
The authors in \cite{Yu_ris_secrecy_globecom,Yu_ris_secrecy,Cui_RIS_secrecy} have studied various optimization techniques to maximize the secrecy rate of RIS enabled wireless systems and designed the jointly optimal transmit beamformer and RIS phase shift matrix. The above papers consider performance optimization rather than secrecy performance analysis. The secrecy performance of an RIS-aided indoor communication system via secrecy outage probability (SOP) and
average secrecy rate is considered in \cite{Tuan_RIS_secrecy} assuming Rician fading model. The SOP is also derived assuming a Rayleigh fading model in \cite{Renzo_ris_secrecy}. These papers do not consider friendly jamming to enhance the secrecy performance or optimize the SOP performance using power allocation.

Artificial noise (AN) can be added to the received signal by the transmitter or a jammer to degrade the signal-to-noise ratio (SNR) at the eavesdropper without affecting the SNR at the destination \cite{rohit_artificial_noise}. 
The legitimate user and the transmitter/jammer cooperate in the system, and thus the user can
cancel out the effect of jamming. Furthermore, cooperative jamming (CJ), where a relay node transmits a jamming signal to create interference at the eavesdropper, is another way to improve the secrecy performance \cite{Cooperative_jamming_poor}. 
The authors in \cite{chinmoy_cooperative_jamming} derived the SOP of a joint relay and friendly jammer selection scheme from a set of multiple energy harvesting relays. 
In \cite{Cumanan_power_allocation_jamming}, the authors considered optimal power allocation to cooperative jammers to maximize the secrecy rate and obtained the corresponding SOP performance .


Though using AN and CJ to improve secrecy is common in the secrecy literature, the application of AN or CJ to RIS-aided system is yet in the nascent stage. The authors in \cite{Guan_Zhang_IRS_artificial_noise} first investigated the joint
transmit beamforming with AN and RIS reflect beamforming
in an RIS-aided secure communication system to maximize the achievable secrecy rate.
The authors in \cite{Wang_cooperative_jamming_ris} obtained the optimal secure beamforming and phase shift matrix for the RIS in a multiple-input single-output (MISO) system  with a friendly jammer by maximizing energy efficiency. 
In \cite{Xu_Cao_irs_empowered_PLS}, the authors deployed an RIS not to aid information transmission, however but as a jammer to redirect the energy of the confidential signal and deteriorate the reception at the eavesdropper. In this context, we also mention that authors in \cite{Sun_Chatzinotas_IRS_enhanced_secure} used jamming to aid eavesdropping and cooperation between the jammer and eavesdropper was considered.


The literature mentioned above with friendly jammers considered the joint design of transmit beamforming and RIS phase shift matrix, to maximize the secrecy rate or energy efficiency. The SOP performance was however not evaluated.  To the best of our knowledge, the SOP analysis using friendly jamming and optimal power allocation to minimize the SOP has not been considered yet in the literature.  
Motivated by this, we consider an RIS-aided system in which the signals from the jammer and the source are impinging on the RIS before being reflected to the destination, and obtain the approximate closed-form SOP performance. Further, power is optimally allocated among the source and the jammer in order to achieve minimum SOP assuming a 
total power constraint. 
\vspace{-0.2cm}
\subsection{Key Contributions}
The main contributions of this work are listed as follows:
\begin{enumerate}
    \item We find the approximate closed-form SOP of an RIS enabled system with a single source, destination, eavesdropper, and a friendly jammer transmitting noise to degrade eavesdropping. 
    \item A simpler approximate closed-form SOP expression is further derived to implement the optimal power allocation at the transmitter and the jammer, and to provide a better insight regarding the performance depends on the system parameters.
    \item Optimal power allocation for the transmitter and the jammer is carried out by proving the convexity of the approximate SOP expression assuming a total power constraint. 
    \item The effect of the various system parameters on the SOP and the optimal power allocation factor is shown.
\end{enumerate}

\textit{Notation:} For a random variable (r.v.) $X$, $\mathbb{E}_X[\cdot]$, $F_{X} (\cdot)$, and $f_{X} (\cdot)$ denote the expectation, cumulative distribution function (CDF), and probability density function (PDF), respectively. $\mathbb{P}[\cdot]$ represents the probability of occurrence of an event.
\section{System and channel model}\label{section_system}
We consider a wireless communication system wherein a source $S$ intends to communicate with a legitimate destination $D$ with the aid of an RIS $R$. A passive eavesdropper $E$ attempts to overhear the communication between $S$ and $D$. The RIS with $N$ elements enables the communication by intelligently reflecting the signals from $S$ towards $D$.
Since $E$ can overhear the communication via RIS, a friendly jammer $J$ is introduced to transmit a jamming signal via $R$ independent of the source signal, to degrade the signal quality at $E$. A scenario is considered where an obstacle is present and the blockage is so severe that the signal from $S$ and $J$ cannot reach $D$ and $E$ directly. The jamming signal is assumed to be known to $D$ and hence, it causes controlled interference at $E$ but not at $D$, thereby enhancing the system's secrecy performance. Nodes $S$, $J$, $D$, and $E$, are assumed to be equipped with a single antenna. The signals from $S$ and $J$ are received by $D$ and $E$ after being reflected from each element of the RIS having amplitude reflection coefficient $\eta_n$ and adjustable phase $\theta_n\in(0,2\pi]$, where $n\in\{1,\ldots,N\}$. 
Transmitting nodes, $S$ and $J$, have a total power budget of $P_T$, of which a fraction $\alpha$ is used by $S$ and a fraction $(1-\alpha)$ fraction by $J$, where $0<\alpha<1$ is the power allocation factor for $S$.
{The channel for node pair combinations   $XY\in\{SR,RD,JR,RE\}$ is denoted as $h_{XY}^{(n)}= \sqrt{\zeta_{XY}}\tilde{h}_{XY}^{(n)}$,  wherein $\tilde{h}_{XY}^{(n)}$ is a complex Gaussian r.v. with zero mean and unit variance and the path loss is $\zeta_{XY} = \big(\frac{d_0 }{d_{XY}}\big)^{\upsilon}$ where  $d_0$ is reference distance, $d_{XY}$ denotes the distance between the nodes, and the path loss exponent $\upsilon$. 
}

The received signal at $D$ through the RIS is given by
\begin{align}\label{signal_user}
y_D&=\sqrt{\alpha P_T}\sum_{n=1}^N h_{RD}^{(n)}\eta_ne^{j\theta_n}h_{SR}^{(n)}
s\nn\\
&+\sqrt{(1-\alpha)P_T}\sum_{n=1}^N h_{RD}^{(n)}\eta_ne^{j\theta_n}h_{JR}^{(n)}z +w_D,
\end{align}
where $s$ and $z$ denote the information and jamming signals transmitted by $S$ and $J$, respectively, with $\mathbb{E}\{|s|^2\}=\mathbb{E}\{|z|^2\}=1$,  $w_D$ denotes the complex additive white Gaussian noise (AWGN) at $D$ with zero-mean and variance $N_0$. 
The received signal at $E$ through the RIS is given by an expression similar to (\ref{signal_user}) but  with $D$ replaced by $E$. Without loss of generality, the noise at $E$ is also modelled as complex AWGN with the same mean and a variance of $w_D$.

\subsection{SINR distribution of the legitimate link}
Assuming RIS phase shift $\theta_n$ for each $n\in\{1,\ldots,N\}$ is adjusted to maximize the SNR at $D$ by constructive signal combining with the knowledge of the phases of $h_{RD}^{(n)}$ and $h_{SR}^{(n)}$; and $D$ has all the knowledge to cancel the interference from $J$, the SINR at $D$ is written as 
\begin{align}\label{snr_user_without_jamming}
\Gamma_D&=\alpha\Gamma_0\Bigg(\sum_{n=1}^N \lb|h_{RD}^{(n)}\rb|\lb|h_{SR}^{(n)}\rb|\Bigg)^2,
\end{align}
where $\Gamma_0={P_T}/{N_0}$ and $\eta_n=1$ for each $n$ is assumed throughout the paper.
Since an RIS typically has large $N$, from the central limit theorem, the distribution of the sum of 
r.v.s within the parenthesis in (\ref{snr_user_without_jamming}) follows a Gaussian distribution, the mean $\mu$ and variance $\sigma^2$ of which is computed as  $\mu=\frac{\pi N
\sqrt{\zeta_{RD}\zeta_{SR}}}{4}$ and $\sigma^2= N
\zeta_{RD}\zeta_{SR}\lb(\frac{16-\pi^2}{16}\rb)$  \cite{Salo_product_of_rayleigh}. 
Therefore, the CDF of $\Gamma_{D}$ is written as \cite{RIS_Amarasuriya}
\begin{align}\label{cdf_gammaD}
F_{\Gamma_D}(x)
&=1-\psi{Q}\Bigg(\frac{\sqrt{\frac{x}{\alpha \Gamma_0}}-\mu}{\sigma}\Bigg) \text{for}~ x\ge0,
\end{align}
where ${Q}(\cdot)$ denote the Gaussian ${Q}$-function and {$\psi=1/Q\lb(-{\mu/\sigma}\rb)$ is the normalization factor.}
\subsection{SINR distribution of the eavesdropping link}
As $E$ cannot cancel out the jamming signal, the SINR at $E$ is expressed as
\begin{align}
\Gamma_E&=\frac{\lb|\sqrt{\alpha P_T}\sum_{n=1}^N h_{RE}^{(n)}\eta_ne^{j\theta_n}h_{SR}^{(n)}\rb|^2}{N_0+\lb|\sqrt{(1-\alpha)P_T}\sum_{n=1}^N h_{RE}^{(n)}\eta_ne^{j\theta_n}h_{JR}^{(n)}\rb|^2}\nn\\
&=\frac{\lb|H_{SE}\rb|^2}{1+\lb|H_{JE}\rb|^2},
\end{align}

where $H_{SE}=\sqrt{\alpha \Gamma_0}\sum_{n=1}^N h_{RE}^{(n)}\eta_ne^{j\theta_n}h_{SR}^{(n)}$ and $H_{JE}=\sqrt{(1-\alpha)\Gamma_0}\sum_{n=1}^N h_{RE}^{(n)}\eta_ne^{j\theta_n}h_{JR}^{(n)}$.

As the RIS phase shift $\theta_n$ is adjusted to benefit $D$, the phases of channels $h_{SR}^{(n)}$ and  $h_{RE}^{(n)}$ for each $n\in\{0,\ldots,N\}$ 
will be misaligned. 
Similarly, the phases of $h_{JR}^{(n)}$ and $h_{RE}^{(n)}$  for each $n\in\{0,\ldots,N\}$ will also be misaligned. Consequently,  $H_{SE}$ and $H_{JE}$ become complex Gaussian r.v.s, and
$|H_{SE}|^2$ and $|H_{JE}|^2$ become exponential r.v.s with parameters   ${\lambda_{SE}}=\alpha\Gamma_0N\eta^2\zeta_{RE}\zeta_{SR}$ and  $\lambda_{JE}=(1-\alpha)\Gamma_0N\eta^2\zeta_{RE}\zeta_{JR}$ , respectively, following \cite{Renzo_ris_secrecy}.
The  PDF of $|H_{XY}|^2$ is thus expressed as
\begin{align}\label{pdf_h_jre}
f_{|H_{XY}|^2}(x)&=\frac{1}{\lambda_{XY}} \exp{\lb(-\frac{x}{\lambda_{XY}}\rb)}. 
\end{align}
Therefore, the  distribution of $\Gamma_E$ is derived 
with the help of  (\ref{pdf_h_jre})  and its CDF as
\begin{align}\label{cdf_gamma_E}
&F_{\Gamma_E}(x)
=\int_0^\infty F_{|H_{SE}|^2}\lb(x{\lb( 1+y\rb)}\rb)f_{|H_{JE}|^2}(y)dy\nn\\
&=
\frac{1}{\lambda_{JE}}\int_0^\infty\lb(
1-\exp{\lb(-\frac{(1+y)x}{\lambda_{SE}}\rb)} \rb)\exp{\lb(-\frac{y}{\lambda_{JE}}\rb)}dy \nn\\
&=1-\frac{\frac{\lambda_{SE}}{\lambda_{JE}}\exp{\lb(-\frac{x}{\lambda_{SE}}\rb)}}{\lb({x+\frac{\lambda_{SE}}{\lambda_{JE}}}\rb)}.
\end{align}

By differentiating (\ref{cdf_gamma_E}), the PDF of $\Gamma_E$ is expressed as
\begin{align}\label{pdf_gammaE}
f_{\Gamma_E}(x)=\frac{\exp{\lb(-\frac{x}{\lambda_{SE}}\rb)}}{\lambda_{JE}\lb({x+\frac{\lambda_{SE}}{\lambda_{JE}}}\rb)}+\frac{{\lambda_{SE}}\exp{\lb(-\frac{x}{\lambda_{SE}}\rb)}}{{\lambda_{JE}}\lb({x+\frac{\lambda_{SE}}{\lambda_{JE}}}\rb)^2}.
\end{align}

\section{Secrecy Outage Probability}\label{section_sop}
This section evaluates the SOP of the system. The instantaneous secrecy rate $C_S$ is defined as the difference between the legitimate channel capacity and the wiretap channel capacity and is expressed as \cite{chinmoy_wcl}
\begin{align}\label{secrecy_capacity}
C_S=\max\lb\{\log_2\lb(\frac{{1+{\Gamma}_{D}}}{1+{\Gamma}_{E}}\rb),0\rb\}. 
\end{align}
The SOP is then defined as the probability that the secrecy rate $C_S$ falls below a threshold rate  $R_{th}$ and is expressed as \cite{chinmoy_twc}
\begin{align}\label{secrecy_ outage_eq}
\mathcal{P}_{out}&= \mathbb P\lb[C_{S} < R_{th}\rb]
\nn \\
&= \int_{0}^{\infty} F_{{\Gamma}_{D}}(\rho(x+1)-1) f_{{\Gamma}_{E}}(x) dx,
\end{align}
where $\rho=2^{R_{th}}$. 
Using $F_{{\Gamma}_{D}}(x)$ and $f_{{\Gamma}_{E}}(x)$  derived in (\ref{cdf_gammaD}) and (\ref{pdf_gammaE}), respectively, in the previous section, the SOP is expressed as
\begin{align}\label{P_out_Q_func}
\mathcal{P}_{out}&= 1-\int_{0}^{\infty} \psi{Q}\lb(\frac{\sqrt{\frac{\rho(x+1)-1}{\alpha\Gamma_0}}-\mu}{\sigma}\rb)f_{{\Gamma}_{E}}(x)dx.
\end{align}
To obtain the SOP in (\ref{P_out_Q_func}), we have to average the $Q$-function which is not trivial. To obtain the integral solution in closed form,  we use the following approximation of $Q$-function \cite{Q_func_approx}: 
\begin{subequations}
  \begin{empheq}[left=Q(x)\\
  \approx\empheqlbrace]{align}
  \label{eq_Q_func_sum_3exp1}
    & 1-\sum_{i=1}^3\frac{ w_i}{2}\exp\lb(-\frac{p_i x^2}{2}\rb) \textrm{when} ~~x<0, \\
    \label{eq_Q_func_sum_3exp2}
    & \sum_{i=1}^3\frac{ w_i}{2}\exp\lb(-\frac{p_i x^2}{2}\rb) ~~~~\textrm{when} ~~x\ge0,
  \end{empheq}
\end{subequations}
where
$w_i=\lb\{\frac{1}{6}, \frac{1}{3}, \frac{1}{3}\rb\} ~\textrm{and}~ p_i=\lb\{1, 4, \frac{4}{3}\rb\}$ for $i\in\{1,2,3\}$.
The integral in (\ref{P_out_Q_func}) is derived based on the positive or negative argument of the $Q$-function following (\ref{eq_Q_func_sum_3exp1}) and (\ref{eq_Q_func_sum_3exp2}) as
 \begin{align}\label{P_out_exponential}
     \mathcal{P}_{out}=1-\lb(\mathcal{P}_{out}^-+\mathcal{P}_{out}^+\rb),
 \end{align}
where 
\begin{align}\label{P_out^-}
\mathcal{P}_{out}^-&=\psi\int_{0}^{a}\lb(1-\sum_{i=1}^3\frac{w_i}{2}\chi(x)\rb) f_{{\Gamma}_{E}}(x)dx,\\
\label{P_out^+}
\mathcal{P}_{out}^+&= \sum_{i=1}^3\frac{ \psi w_i}{2}\int_{a}^{\infty}\chi(x)f_{{\Gamma}_{E}}(x)dx,
\end{align}
and  $\chi(x)=\exp{\Big(-\frac{p_i}{2\sigma^2}\big(\sqrt{\frac{\rho(x+1)-1}{\alpha\Gamma_0}}-\mu\big)^2\Big)}$. 
The value $a$ which serves as the upper and lower limit of the integrals in (\ref{P_out^-}) and (\ref{P_out^+}), respectively, is evaluated by finding limits of $x$ 
for which $\frac{1}{\sigma}\Big(\sqrt{\frac{\rho(x+1)-1}{\alpha \Gamma_0}}-\mu\Big)$ is greater or less than zero. Thus, we obtain 
 $a=\frac{\alpha\mu^2\Gamma_0+1}{\rho}-1$.

Here we note that $\mathcal{P}_{out}^+$ tends to zero as $a$ becomes large in (\ref{P_out^+}) when $\Gamma_0$ or $N$ or both are large. Thus, the effect of $\mathcal{P}_{out}^+$ is negligible as compared to $\mathcal{P}_{out}^-$ and the SOP approximates to $\mathcal{P}_{out}\approx1-\mathcal{P}_{out}^-$.
Then $\mathcal{P}_{out}^-$ is  evaluated as
\begin{align}\label{eq_P_out_minus}
\mathcal{P}_{out}^-=I_0-I_1-I_2,
\end{align}
where
\begin{align}\label{I_0}
I_{0}&=\psi\int_{0}^{a}f_{\Gamma_{E}}(x)dx\nn\\
&=\psi\Bigg[\frac{e^{\frac{1}{\lambda_{JE}}}}{\lambda_{JE}}\left(\text{Ei}\left(-\frac{a}{\lambda_{SE}}-\frac{1}{\lambda_{JE}}\right)-\text{Ei}\left(-\frac{1}{\lambda_{JE}}\right)\right.\Bigg.\nn\\
&\Bigg.\left.+\Gamma \left(-1,\frac{1}{\lambda_{JE}}\right)-\Gamma \left(-1,\frac{a}{\lambda_{SE}}+\frac{1}{\lambda_{JE}}\right)\right)\Bigg],\\
\label{I_1}
{I}_{1}&=\sum_{i=1}^3\frac{\psi w_i}{2}\int_{0}^{a}\chi(x)\frac{\exp{\lb(-\frac{x}{\lambda_{SE}}\rb)}}{\lambda_{JE}\lb({x+\frac{\lambda_{SE}}{\lambda_{JE}}}\rb)}dx,\\
\label{I_2}
{I}_{2}&=\sum_{i=1}^3\frac{\psi w_i}{2}\int_{0}^{a}\chi(x)\frac{{\lambda_{SE}}\exp{\lb(-\frac{x}{\lambda_{SE}}\rb)}}{{\lambda_{JE}}\lb({x+\frac{\lambda_{SE}}{\lambda_{JE}}}\rb)^2}dx.
\end{align}
The solutions of the integrals in (\ref{I_1}) and (\ref{I_2}) are similar. 
After some algebraic manipulations, $I_{1}$ is expressed as 
\begin{align}\label{eq_I_1}
{I}_{1}&=\sum_{i=1}^3\frac{\psi w_i\xi}{2\lambda_{JE}}\Psi,
\end{align}
where 
\begin{align}
\xi&=\exp\lb[-\frac{p_i\mu^2 }{2 \sigma ^2}+\frac{p_i(\rho -1) \lambda_{SE}\Gamma_0}{ \rho p_i\lambda_{SE} +2\alpha\sigma^2\Gamma_0}+\frac{ \rho p_i\lambda_{SE}+2 \alpha\sigma ^2\Gamma_0 }{2\alpha \sigma^2\lambda_{SE}\Gamma_0 }\rb.\nn\\
&\lb.\times\left(\frac{\alpha\rho p_i^2\mu^2\lambda_{SE}^2\Gamma_0}{\left( \rho p_i\lambda_{SE}+2\alpha\sigma^2\Gamma_0\right)^2}+\frac{\rho -1}{\rho }\right)\rb],    
\end{align}
and
\begin{align}
\label{eq_integral_psi}
\Psi&=\int_{\sqrt{\rho-1}}^{\sqrt{a\rho+\rho-1}}\frac{x\exp{\lb(-b\lb(x-c\rb)^2\rb)}}{x^2+d}dx,
\end{align}
with $b=\frac{ \rho p_{i} \zeta_{RE} \zeta_{SR} +2\sigma^2}{2 \alpha \rho \zeta_{RE} \zeta_{SR}\sigma^2 \Gamma_0 N }$, $c=\frac{\sqrt{\alpha \Gamma_0}\rho p_{i}\mu \zeta_{RE} \zeta_{SR}}{\rho p_{i} \zeta_{RE} \zeta_{SR}+2\sigma^2}$, and $d=1-\rho +\frac{\lambda_{SE}}{\lambda_{JE}}$. The solution of (\ref{eq_integral_psi}) is difficult to achieve, hence, we find an approximation of the same. 
By noting that  $\exp{\big(-b\lb(x-c\rb)^2\big)}$ is symmetric around $x=c$ and if $c$ is large, which is typically the case when $N$ is large or $\Gamma_0$ is large (high-SNR operation) or both, the product (\ref{eq_integral_psi}) tends to be symmetric around $x=c$, we replace $\frac{x}{x^2+d}$ in (\ref{eq_integral_psi}) by 
its first order Taylor series approximation around $x=c$, i.e.,
$\frac{x}{x^2+d}\approx\frac{c}{c^2+d}$.
Then, $\Psi$ is easily approximated as
\begin{align}\label{Psi}
&\Psi=
\frac{c\sqrt{\pi}}{2 \sqrt{b}(c^2+d)}  \left[\text{erf}\left(\sqrt{b} \left(\sqrt{a\rho+\rho-1}-c\right)\right)\right.\nn\\
&\left.-\text{erf}\left(\sqrt{b} \left(\sqrt{\rho-1}-c\right)\right)\right].
\end{align}

In the evaluation of $I_2$, we follow the same steps as were used while evaluating $I_{1}$  and use the approximation  $\frac{x}{x^2+d^2}\approx\frac{c}{(c^2+d)^2}$. 
Therefore, $I_{2}$ is evaluated as
\begin{align}\label{I_2_closed_form}
&I_{2}=\sum_{i=1}^3\frac{\psi c\sqrt{\pi} w_i\xi\lambda_{SE}}{2 \sqrt{b}(c^2+d)^2\lambda_{JE}}\left[\text{erf}\left(\sqrt{b} \left(\sqrt{a\rho+\rho-1}-c\right)\right)\right.\nn\\
&-\left.\text{erf}\left(\sqrt{b} \left(\sqrt{\rho-1}-c\right)\right)\right].
\end{align}

We observe from (\ref{eq_P_out_minus}), (\ref{I_0}), (\ref{eq_I_1}) and (\ref{I_2_closed_form}) that the closed-form approximate SOP does not provide adequate information about how it behaves with the different system parameters  $\alpha$, $\rho$, $N$, $\Gamma_0$ and  $\zeta_{XY}$ where $XY\in\{SR,RD,JR,RE\}$.
To obtain better insights into the SOP performance, we use a further approximation in the following section. This will also help us to obtain the optimal power allocation factor $\alpha$ in the next section.
\section{Optimal Power Allocation}\label{section_optimal}

In this section, the optimal power allocation factor, $\alpha^*$, is derived which minimizes the approximate SOP. To find $\alpha^*$, we first prove the convexity of the approximate SOP with respect to $\alpha$. This is difficult to prove from the approximate SOP derived in (\ref{P_out_exponential}) in the previous section; thus,
we use further approximations on the SOP assuming $N$ is large. The approximate SOP in this section also provides insights about the system behaviour. 
The integrals in (\ref{I_0}), (\ref{I_1}), and (\ref{I_2}) are respectively evaluated by assuming $a\rightarrow\infty$ when $N$ is large following the same approach considered in the previous section
as
\begin{align}
\label{eq_I012_high}
I_{0}&=\psi,
I_{1}=\psi\sum_{i=1}^3\frac{ w_ic\rho\xi\sqrt{\pi}}{2 \sqrt{b}(c^2+d)\lambda_{JE}},
\nn\\
I_{2}&=\sum_{i=1}^3\frac{\psi w_ic\xi\sqrt{\pi}\lambda_{SE}}{2 \sqrt{b}(c^2+d)^2\lambda_{JE}}.
\end{align}
After using $I_0$, $I_1$, and $I_2$ from (\ref{eq_I012_high}) in (\ref{eq_P_out_minus}) and substituting for the values of $b$, $c$, $\lambda_{SE}$ and $\lambda_{JE}$, the approximate SOP  is written assuming $R_{th}>0$, i.e., $\rho>1$, as
\begin{align}
\label{eq_SOP_high}
\mathcal{P}_{out}&=\sum_{i=1}^3\frac{\Omega_1 \Omega_2}{1-\alpha}\lb(1+\frac{\rho\lambda_{SE}}{c^2+d}\rb)\exp\left(\frac{\rho-1}{\alpha\rho \zeta_{RE} \zeta_{SR}\Gamma_0N }\right),
\end{align}
where
\begin{align}\label{omega_1}
&\Omega_1=\frac{\psi w_{i}\sqrt{\lb({16-\pi^2}\rb)}}
{2\rho p_{i}{N^{3/2}}\sqrt{\pi} \Gamma_0\zeta_{JR}\zeta_{RE}}\sqrt{{2\rho^2 p_{i}+ \frac{(16-\pi^2)\rho\zeta_{RD}}{4\zeta_{RE}}}}\\\label{omega_2}
&\Omega_2=\exp\left[-\frac{p_{i}\pi^2N }{2\lb({16-\pi^2}\rb)}
\lb(\frac{{1}}
{\frac{8\rho p_{i}\zeta_{RE}}{(16-\pi^2)\zeta_{RD}}+{1}}\rb)\right].
\end{align}

To further simplify the equation, the term $\frac{\rho\lambda_{SE}}{(c^2+d)}$ is neglected in (\ref{eq_SOP_high}) as $\frac{\rho\lambda_{SE}}{(c^2+d)}<<1$ when $N$ is large. Finally, the approximate SOP in expressed in a compact form as a function of $\alpha$ as
\begin{align}\label{p_out_approximate}
&\mathcal{P}_{out}=\sum_{i=1}^3\frac{\Omega_1 \Omega_2}{1-\alpha}\exp\left(\frac{\rho-1}{\alpha\rho \zeta_{RE} \zeta_{SR}\Gamma_0N }\right).
\end{align}

Now, it is easy to understand how the SOP behaves with the system parameters $\alpha$,  $N$, $\Gamma_0$ and  $\zeta_{XY}$ where $XY\in\{SR,RD,JR,RE\}$. We observe from (\ref{p_out_approximate}) along with (\ref{omega_1}) and (\ref{omega_2}) that the SOP is inversely proportional to $\Gamma_0$. This is because $\Omega_1$ is inversely proportional to $\Gamma_0$ and the exponential term in (\ref{p_out_approximate})  tends to unity as $N$ becomes large. From (\ref{omega_1}) and (\ref{omega_2}) we observe that the SOP decreases with $N$ at a rate of $\exp(-N)/N^{3/2}$ as the exponential term in (\ref{p_out_approximate})  tends to unity as $N$ increases. 
As the distance between the RIS and the eavesdropper $d_{RE}$ increases, the SOP decreases exponentially. This is because $\zeta_{RE}$ is inside the exponential in (\ref{omega_2}) and  dominates over (\ref{omega_1}) and (\ref{p_out_approximate}) affecting the SOP. 
As the distance between the RIS and the destination $d_{RD}$ increases, from (\ref{omega_2}) we observe that the SOP increases exponentially and it dominates (\ref{omega_1}) which decreases slowly.  
As the distance between the jammer and the RIS  $d_{JR}$ increases, the SOP increases as we observe from (\ref{omega_1}).
As the distance between the transmitter and the RIS $d_{SR}$ changes, the SOP does not change as $\Omega_1$ and $\Omega_2$ are independent of $\zeta_{SR}$. Also, although $\zeta_{SR}$ is present in (\ref{p_out_approximate}), it does not affect the SOP as the exponential term in (\ref{p_out_approximate}) tends to unity as $N$ increases.

Finally, we observe from (\ref{p_out_approximate}) that as $\alpha$ increases, the exponential term in (\ref{p_out_approximate}) decreases and hence, the SOP decreases. In contrast, the denominator $(1-\alpha)$ decreases and hence, the SOP increases. This suggests that there is an optimal value of $\alpha$ which can minimize the SOP. The optimal $\alpha$ is derived in the next section.


\subsection{Finding the optimal power allocation}
To check the convexity of the approximate SOP with respect to $\alpha$, we need to double differentiate the approximate SOP equation in (\ref{p_out_approximate}). 
The first derivative of $\mathcal{P}_{out}$ is given by
\begin{align}\label{first_derivative}
\frac{d \mathcal{P}_{out}}{d\alpha}&=\sum_{i=1}^3 \frac{\Omega_1 \Omega_2 \exp \lb({\frac{\rho -1}{\alpha  \rho \zeta_{RE} \zeta_{SR} \Gamma_0 N}}\rb)}{1-\alpha}\nn\\
&\times\lb(\frac{1}{1-\alpha }-\frac{\rho -1}{\alpha^2 \rho \zeta_{RE} \zeta_{SR}\Gamma_0 N }\rb),
\end{align}
and the second derivative is evaluated as 
 \begin{align}\label{second_derivative}
&\frac{d^2 \mathcal{P}_{out}}{d\alpha^2}
=\sum_{i=1}^3\frac{\Omega_1\Omega_2 \exp\lb({\frac{\rho-1}{\alpha\rho \zeta_{RE} \zeta_{SR}\Gamma_0N }}\rb)}{1-\alpha }\lb(\frac{2-(1-\alpha )^2}{(1-\alpha )^2}\rb.\nn\\
&\lb.+\frac{(\rho -1)^2}{\alpha^4 \rho^2 \zeta _{RE}^2 \zeta_{SR}^2 N^2 \Gamma_0^2}+\frac{2 (\rho -1)}{(1-\alpha)\alpha^3 \rho \zeta_{RE}\zeta_{SR} \Gamma_0 N}\rb).
\end{align}
We observe that since the parameters $\rho$, $\zeta_{RE}$, $\zeta_{SR}$, $N$, and $\Gamma_0$ are all positive and $0<\alpha<1$ and $\rho>1$,
(\ref{second_derivative}) is positive for all possible values of $\alpha$ and hence the function is convex with respect to $\alpha$ and a global optimal minimum value exists.

Now, equating the first derivative of $\mathcal{P}_{out}$ in (\ref{first_derivative}) to zero, we can find $\alpha^*$. The expression in (\ref{first_derivative}) will be equal to zero when the term inside the brackets is equal to zero. After some mathematical manipulations, this condition reduces to the quadratic equation as 
\begin{align}
\alpha ^2 \rho\zeta_{RE} \zeta_{SR} \Gamma_0 N+\alpha  (\rho -1)-(\rho-1)=0. 
\end{align}
Since $\alpha^*$ cannot be a negative quantity, we take the positive root for $\alpha$ and therefore the optimal power allocation factor is given by
\begin{align}\label{optimal_alpha}
\alpha^*=\frac{-(\rho -1)+\sqrt{(\rho -1)^2+4 \rho\zeta_{RE} \zeta_{SR}\Gamma_0 N (\rho-1 )}}{2 \rho\zeta_{RE} \zeta_{SR}  \Gamma_0 N}.
\end{align}
We observe from (\ref{optimal_alpha}) that the optimal power allocation depends on parameters $\rho$, $\zeta_{SR}$ and $\zeta_{RE}$, $N$ and $\Gamma_0$. However, it is independent of  $\zeta_{JR}$ and $\zeta_{RD}$. As $\rho$, $\zeta_{SR}$ and $\zeta_{RE}$, $N$, and $\Gamma_0$ increase, the jamming power needs to be increased in order to minimize the SOP.
\section{Numerical Results and Discussion}\label{section_results}

\begin{figure}
    \centering
    \includegraphics[width=3.081in]{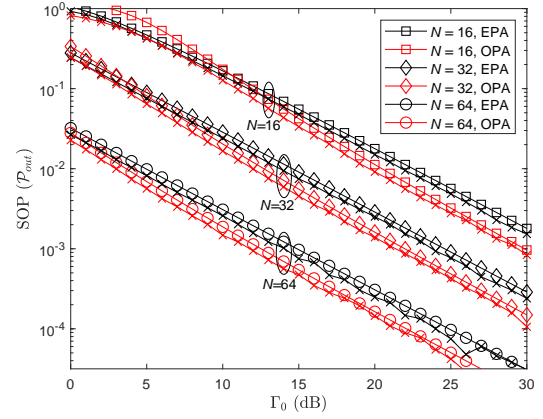}
    \vspace{-.24cm}
    \caption{SOP versus $\Gamma_0$ for different values of $N$.
    }
    \label{sop1_versus_snr}
    \vspace{-.55cm}
\end{figure}

In this section, analytical results are plotted along with the numerical results. Numerical results are denoted by a solid line with the symbol `$\times$' having the same color of the corresponding analytical curve.
 The system parameters, unless otherwise stated, are considered as $R_{th}=1$ bit per channel use, $\{d_{SR},d_{JR},d_{RD},d_{RE}\} = \{30, 30, 30, 15\}$ m, the path loss in dB is given by $\zeta_{XY}(\text{dB}) = z_0-10v\log_{10}(d_{XY})$ where
 {$z_0=42$} dB and  $v=3.5$ is the reference path loss and path loss factor, respectively. 

Fig. \ref{sop1_versus_snr} plots the approximate SOP from (\ref{P_out_exponential}) 
versus $\Gamma_0$ for different number of RIS elements $N\in\{16, 32, 64\}$ with equal power allocation (EPA) $\alpha=0.5$ and with the optimal power allocation (OPA) $\alpha^*$. We find that the SOP decreases with the increase in the number of RIS elements. It is observed that the OPA outperforms the EPA as expected. By using the OPA factor, a gain of approximately $3$ dB is achieved at an SOP of $10^{-4}$ for $N=64$.




\begin{figure}
    \centering
    \includegraphics[width=3.081in]{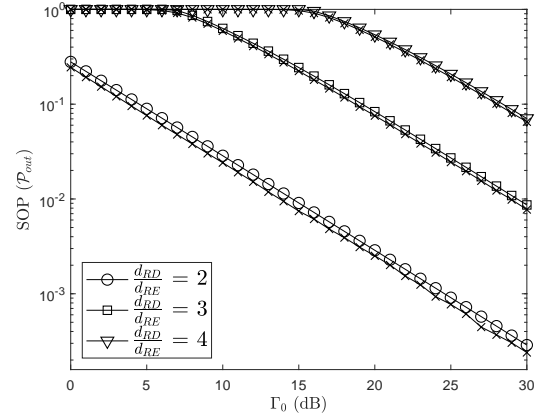}
    \vspace{-.24cm}
    \caption{SOP versus $\Gamma_0$ for different values of $d_{RD}/d_{RE}$.
    }
    \label{sop_v_snr_distance_ratio}
    \vspace{-.55cm}
\end{figure}

In Fig. \ref{sop_v_snr_distance_ratio}, the approximate SOP from (\ref{P_out_exponential})  is plotted versus $\Gamma_0$, by varying the ratio of distances $R$-$D$ and $R$-$E$, i.e., $\frac{d_{RD}}{d_{RE}}$ with $\{d_{SR},d_{JR}, d_{RE}\}$ = $\{30, 30, 15\}$ m. It is observed that as the ratio increases, the secrecy performance decreases. 
    This signifies that as the $R$-$D$ distance increases, or as the $R$-$E$ distance decreases, the SOP increases.

\begin{figure}
    \centering
    \includegraphics[width=3.081in]{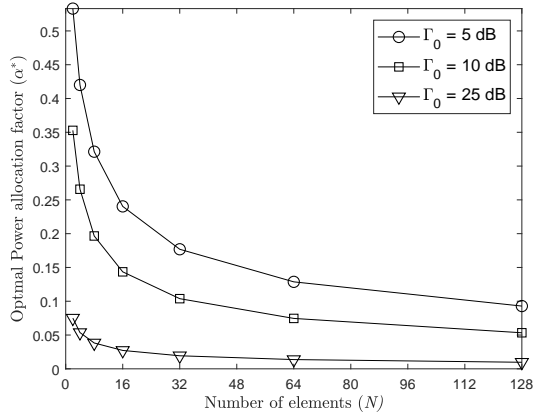}
    \vspace{-.24cm}
    \caption{Optimal $\alpha^*$ for different numbers of RIS elements $N$.}
    \label{N versus optimal alpha}
    \vspace{-.55cm}
\end{figure}

Fig. \ref{N versus optimal alpha} shows the effect of $N$ on $\alpha^*$ at different $\Gamma_0$. It is observed that as $N$ increases, $\alpha^*$ decreases, i.e., more power is allocated to the jammer. This is due to the fact that as the number of RIS elements increases, the reflected signal strength increases for the eavesdropper as well. The jamming power has to be increased in order to minimize the SOP. We also notice that as $\Gamma_0$ increases for a given $N$,  $\alpha^*$ decreases. This means that as the SNR improves, less power is required for signal transmission to minimize the SOP.  

\begin{figure}
    \centering
    \includegraphics[width=3.081in]{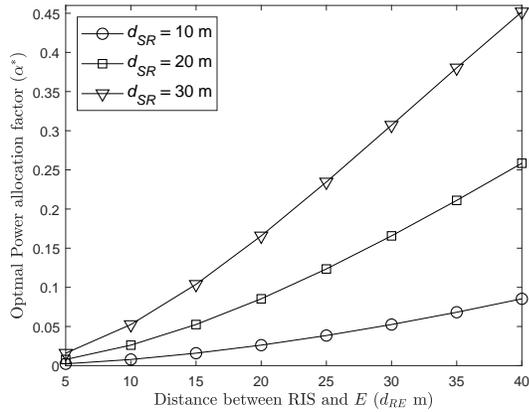}
    \vspace{-.24cm}
    \caption{Optimal $\alpha^*$ versus $d_{RE}$ for different locations of $E$.}
    \label{dREalpha}
    \vspace{-.55cm}
\end{figure}

In Fig. \ref{dREalpha},  $\alpha^*$ versus $d_{RE}$ is plotted for different $d_{SR}$ values.  With the increase in $d_{RE}$ for a given $d_{SR}$ or with the increase in $d_{SR}$ for a given $d_{RE}$, $\alpha^*$ increases, i.e., more power is allocated for signal transmission. This suggests that as distance of $S$ increases from $R$, more signal  power is required and as distance of $E$ increases from $R$, less jamming power is required to achieve the minimum SOP.


\section{Conclusions}\label{section_conclusions}
The secrecy performance of an RIS-assisted wireless communication has been analyzed in the presence of an eavesdropper. A friendly jammer is introduced to enhance the secrecy performance of the system. The approximate closed-form SOP of the system is evaluated. A simpler approximate closed-form SOP expression is also provided which shows the effect of the system parameters on the performance. This also helps us to show the convexity of the approximate SOP with the power allocation factor. 
The power allocation factor for the transmitter and the jammer is derived in the closed-from.
We find that the optimal power allocation achieves  approximately 3 dB performance gain as compared to the case of equal power allocation at an SOP of $10^{-4}$ for $N=64$.
\bibliographystyle{IEEEtran}
\bibliography{RIS}

\begin{thebibliography}{10}
\providecommand{\url}[1]{#1}
\csname url@samestyle\endcsname
\providecommand{\newblock}{\relax}
\providecommand{\bibinfo}[2]{#2}
\providecommand{\BIBentrySTDinterwordspacing}{\spaceskip=0pt\relax}
\providecommand{\BIBentryALTinterwordstretchfactor}{4}
\providecommand{\BIBentryALTinterwordspacing}{\spaceskip=\fontdimen2\font plus
\BIBentryALTinterwordstretchfactor\fontdimen3\font minus
  \fontdimen4\font\relax}
\providecommand{\BIBforeignlanguage}[2]{{%
\expandafter\ifx\csname l@#1\endcsname\relax
\typeout{** WARNING: IEEEtran.bst: No hyphenation pattern has been}%
\typeout{** loaded for the language `#1'. Using the pattern for}%
\typeout{** the default language instead.}%
\else
\language=\csname l@#1\endcsname
\fi
#2}}
\providecommand{\BIBdecl}{\relax}
\BIBdecl

\bibitem{Zhang_towards_ris}
Q.~Wu and R.~Zhang, ``Towards smart and reconfigurable environment: intelligent
  reflecting surface aided wireless network,'' \emph{IEEE Communications
  Magazine}, vol.~58, no.~1, pp. 106--112, Jan. 2020.

\bibitem{renzo_wireless_com_RIS}
E.~Basar, M.~Di~Renzo, J.~De~Rosny, M.~Debbah, M.-S. Alouini, and R.~Zhang,
  ``Wireless communications through reconfigurable intelligent surfaces,''
  \emph{IEEE Access}, vol.~7, pp. 116\,753--116\,773, Aug. 2019.

\bibitem{Wu_cognitive_ris}
W.~Wu, Z.~Wang, L.~Yuan, F.~Zhou, F.~Lang, B.~Wang, and Q.~Wu, ``{IRS}-enhanced
  energy detection for spectrum sensing in cognitive radio networks,''
  \emph{IEEE Wireless Communications Letters}, vol.~10, no.~10, pp. 2254--2258,
  Oct. 2021.

\bibitem{Xu_user_selection_ris}
X.~Gan, C.~Zhong, Y.~Zhu, and Z.~Zhong, ``User selection in reconfigurable
  intelligent surface assisted communication systems,'' \emph{IEEE
  Communications Letters}, vol.~25, no.~4, pp. 1353--1357, Apr. 2021.

\bibitem{RIS_Amarasuriya}
D.~Kudathanthirige, D.~Gunasinghe, and G.~Amarasuriya, ``Performance analysis
  of intelligent reflective surfaces for wireless communication,'' in
  \emph{Proc. IEEE International Conference on Communications}, Jul. 2020, pp.
  1--6.

\bibitem{wyner_wiretap}
A.~D. Wyner, ``The wire-tap channel,'' \emph{Bell System Technical Journal},
  vol.~54, no.~8, pp. 1355--1387, Oct. 1975.

\bibitem{Yu_ris_secrecy_globecom}
X.~Yu, D.~Xu, and R.~Schober, ``Enabling secure wireless communications via
  intelligent reflecting surfaces,'' in \emph{Proc. IEEE Global Communications
  Conference}, Dec. 2019, pp. 1--6.

\bibitem{Yu_ris_secrecy}
X.~Yu, D.~Xu, Y.~Sun, D.~W.~K. Ng, and R.~Schober, ``Robust and secure wireless
  communications via intelligent reflecting surfaces,'' \emph{IEEE Journal on
  Selected Areas in Communications}, vol.~38, no.~11, pp. 2637--2652, Nov.
  2020.

\bibitem{Cui_RIS_secrecy}
M.~Cui, G.~Zhang, and R.~Zhang, ``Secure wireless communication via intelligent
  reflecting surface,'' \emph{IEEE Wireless Communications Letters}, vol.~8,
  no.~5, pp. 1410--1414, Oct. 2019.

\bibitem{Tuan_RIS_secrecy}
V.~P. Tuan and I.~P. Hong, ``Secrecy performance analysis and optimization of
  intelligent reflecting surface-aided indoor wireless communications,''
  \emph{IEEE Access}, vol.~8, pp. 109\,440--109\,452, Jun. 2020.

\bibitem{Renzo_ris_secrecy}
L.~Yang, J.~Yang, W.~Xie, M.~O. Hasna, T.~Tsiftsis, and M.~D. Renzo, ``Secrecy
  performance analysis of {RIS}-aided wireless communication systems,''
  \emph{IEEE Transactions on Vehicular Technology}, vol.~69, no.~10, pp.
  12\,296--12\,300, Oct. 2020.

\bibitem{rohit_artificial_noise}
R.~Negi and S.~Goel, ``Secret communication using artificial noise,'' in
  \emph{Proc. IEEE Vehicular Technology Conference}, vol.~3, Sep. 2005, pp.
  1906--1910.

\bibitem{Cooperative_jamming_poor}
L.~Dong, Z.~Han, A.~P. Petropulu, and H.~V. Poor, ``Cooperative jamming for
  wireless physical layer security,'' in \emph{Proc. IEEE/SP Workshop on
  Statistical Signal Processing}, Sep. 2009, pp. 417--420.

\bibitem{chinmoy_cooperative_jamming}
T.~M. Hoang, T.~Q. Duong, N.-S. Vo, and C.~Kundu, ``Physical layer security in
  cooperative energy harvesting networks with a friendly jammer,'' \emph{IEEE
  Wireless Communications Letters}, vol.~6, no.~2, pp. 174--177, Apr. 2017.

\bibitem{Cumanan_power_allocation_jamming}
K.~Cumanan, G.~C. Alexandropoulos, Z.~Ding, and G.~K. Karagiannidis, ``Secure
  communications with cooperative jamming: optimal power allocation and secrecy
  outage analysis,'' \emph{IEEE Transactions on Vehicular Technology}, vol.~66,
  no.~8, pp. 7495--7505, Aug. 2017.

\bibitem{Guan_Zhang_IRS_artificial_noise}
X.~Guan, Q.~Wu, and R.~Zhang, ``Intelligent reflecting surface assisted secrecy
  communication: Is artificial noise helpful or not?'' \emph{IEEE Wireless
  Communications Letters}, vol.~9, no.~6, pp. 778--782, Jan. 2020.

\bibitem{Wang_cooperative_jamming_ris}
Q.~Wang, F.~Zhou, R.~Q. Hu, and Y.~Qian, ``Energy efficient robust beamforming
  and cooperative jamming design for {IRS}-assisted {MISO} networks,''
  \emph{IEEE Transactions on Wireless Communications}, vol.~20, no.~4, pp.
  2592--2607, Apr. 2021.

\bibitem{Xu_Cao_irs_empowered_PLS}
S.~Xu, J.~Liu, and Y.~Cao, ``Intelligent reflecting surface empowered
  physical-layer security: Signal cancellation or jamming?'' \emph{IEEE
  Internet of Things Journal}, vol.~9, no.~2, pp. 1265--1275, Jan. 2022.

\bibitem{Sun_Chatzinotas_IRS_enhanced_secure}
Y.~Sun, K.~An, J.~Luo, Y.~Zhu, G.~Zheng, and S.~Chatzinotas, ``Intelligent
  reflecting surface enhanced secure transmission against both jamming and
  eavesdropping attacks,'' \emph{IEEE Transactions on Vehicular Technology},
  vol.~70, no.~10, pp. 11\,017--11\,022, Oct. 2021.

\bibitem{Salo_product_of_rayleigh}
J.~Salo, H.~El-Sallabi, and P.~Vainikainen, ``The distribution of the product
  of independent {R}ayleigh random variables,'' \emph{IEEE Transactions on
  Antennas and Propagation}, vol.~54, no.~2, pp. 639--643, Feb. 2006.

\bibitem{chinmoy_wcl}
C.~Kundu and M.~F. Flanagan, ``Ergodic secrecy rate of optimal source selection
  in a multi-source system with unreliable backhaul,'' \emph{IEEE Wireless
  Communications Letters}, vol.~10, no.~5, pp. 1118--1122, May. 2021.

\bibitem{chinmoy_twc}
C.~Kundu, S.~Ghose, and R.~Bose, ``Secrecy outage of dual-hop regenerative
  multi-relay system with relay selection,'' \emph{IEEE Transactions on
  Wireless Communications}, vol.~14, no.~8, pp. 4614--4625, Aug. 2015.

\bibitem{Q_func_approx}
D.~Sadhwani, R.~N. Yadav, and S.~Aggarwal, ``Tighter {b}ounds on the {G}aussian
  {$Q$} {f}unction and {i}ts {a}pplication in {N}akagami-${m}$ {f}ading
  {c}hannel,'' \emph{IEEE Wireless Communications Letters}, vol.~6, no.~5, pp.
  574--577, Oct. 2017.

\end{thebibliography}
\end{document}